\documentclass{article}
\usepackage{authblk}
\usepackage{geometry}                
\usepackage[parfill]{parskip}        
\usepackage{graphicx}
\usepackage{grffile}
\usepackage{float}
\usepackage{amssymb}
\usepackage{epstopdf}
\usepackage{color}
\usepackage{footmisc}
\usepackage{cite}
\usepackage{url}
\usepackage{amsfonts}
\usepackage{amsmath}
\usepackage{amssymb}
\usepackage{epsfig}
\usepackage{lineno}
\usepackage{hyperref}
\usepackage{subfigure}
\usepackage[ugly]{units}
\usepackage[pscoord]{eso-pic}
\usepackage{geometry}                
\usepackage{graphicx}
\usepackage{float}
\usepackage{amssymb}
\usepackage{epstopdf}
\usepackage{color}
\usepackage[utf8]{inputenc}
\usepackage[normalem]{ulem} 
\usepackage{units}
\usepackage{xcolor}
\usepackage{svn-multi}
\graphicspath{{./}}
\newif\ifcomment
\newif\ifdraft
\newif\ifprefinal
\ifdraft
\linenumbers
\usepackage{fancyhdr}
\fi
\newcommand{\eAu}        {e--Au}

\newcommand{\PbPb}       {Pb--Pb}
\newcommand{\pPb}        {p--Pb}
\newcommand{\pp}         {pp}

\newcommand{\pT}         {\ensuremath{p_{\rm T}}}
\newcommand{\pt}         {\pT}

\newcommand{\GeVc}       {\ensuremath{\mathrm{GeV}/c}}

\newcommand{\Fig}[1]     {Fig.~{#1} in \cite{ALICECollaboration:2020rog}}

\newcommand{\h}[1]       {\relax}

\begin{document}
\title{Snowmass 2021/22 Letter of Interest:\\ A Forward Calorimeter at the LHC}
\author[5]{I.G.~Bearden}
\author[1]{R.~Bellwied}
\author[10]{V.~Borshchov}
\author[12]{J. Faivre}
\author[12]{C.~Furget}
\author[2]{E.~Garcia-Solis}
\author[9]{M.B.~Gay Ducati}
\author[12]{G.~Conesa-Balbastre}
\author[12]{R.~Guernane}
\author[3]{C.~Loizides}
\author[11]{J.~Rojo}
\author[4]{M.~P\l osko\'{n}}
\author[4]{S.R.~Klein}
\author[15]{Y.~Kovchegov}
\author[7]{V.A.~Okorokov}
\author[11]{T.~Peitzmann}
\author[10]{M.~Protsenko}
\author[13]{J.~Putschke}
\author[8]{D.~R\"ohrich}
\author[6]{ J.D.~Tapia Takaki}
\author[10]{I.~Tymchuk}
\author[11]{M.~van Leeuwen}
\author[14]{R.~Venugopalan}
\affil[1]{University of Houston, Houston, Texas, United States}
\affil[2]{Chicago State University, Chicago, Illinois, United States}
\affil[3]{Oak Ridge National Laboratory, Oak Ridge, Tennessee, United States}
\affil[4]{Lawrence Berkeley National Laboratory, Berkeley, California, United States}
\affil[5]{Niels Bohr Institute, University of Copenhagen, Copenhagen, Denmark}
\affil[6]{University of Kansas, Lawrence, Kansas, United States}
\affil[7]{NRNU MEPhI, Moscow, Russia}
\affil[8]{University of Bergen, Bergen, Norway}
\affil[9]{Universidade Federal do Rio Grande do Sul (UFRGS), Porto Alegre, Brazil}
\affil[10]{LTU LLC, Kharkiv, Ukraine}
\affil[11]{Utrecht University/Nikhef, Utrecht, Netherlands}
\affil[12]{LPSC, CNRS-IN2P3, Grenoble, France}
\affil[13]{Wayne State University, Detroit, United States}
\affil[14]{Brookhaven Nation Laboratory, Upton, New York, United States}
\affil[15]{Ohio State University, Columbus, Ohio, United States}
\date{31 Aug 2020}
\maketitle
A forward electromagnetic and hadronic calorimeter (FoCal) was proposed as an upgrade to the ALICE experiment, to be installed during LS3 for data-taking in 2027--2029 at the LHC. 
The FoCal extends the scope of ALICE, which was designed for the comprehensive study of hot and dense partonic matter, by adding new capabilities to explore the small-$x$ parton structure of nucleons and nuclei~\cite{ALICECollaboration:2020rog}.

In particular, the FoCal provides unique capabilities at the LHC to investigate Parton Distribution Functions~(PDFs)~\cite{Kovarik:2015cma,Eskola:2016oht,AbdulKhalek:2019mzd} in the as-yet unexplored regime of Bjorken-$x$ down to $x\sim10^{-6}$  and low momentum transfer $Q\sim4$~\GeVc, where the PDFs are expected to evolve non-linearly~\cite{Balitsky:1995ub,Kovchegov:1999yj,Mueller:2001uk,Lappi:2016fmu} due to the high gluon densities, perhaps leading to {\it saturation}~\cite{Gribov:1984tu,Mueller:1985wy,Ayala:1996em,AyalaFilho:1997du,Kovchegov:1999ua,Iancu:2001md}.
The primary objective of the FoCal is high-precision inclusive measurement of direct photons and jets, as well as coincident gamma-jet and jet-jet measurements, in \pp\ and \pPb\ collisions.
These measurements by FoCal constitute an essential part of a comprehensive small-$x$ program at the LHC down to $x\sim10^{-6}$ and over a large range of $Q^2$ with a broad array of complementary probes, comprising ---in addition to the photon measurements by FoCal and LHCb--- Drell-Yan and open charm measurements planned by LHCb, as well as photon-induced reactions performed by all LHC experiments~\cite{Citron:2018lsq}.
This program will provide by far the most extensive exploration of non-linear effects at small-$x$ for the foreseeable future\h{ until the operation of one of the considered e-h colliders at CERN~(LHeC and FCC-he)}~(\Fig{13}).
Such effects are a necessary consequence of the non-Abelian nature of QCD\h{, but have never been observed}, and their observation and characterization would be a landmark in our understanding of the strong interaction.
Recent small-$x$ computations at next-to-leader order~\cite{Benic:2016uku,Benic:2018hvb,Roy:2019hwr}  will be extended to higher order within the next decade, and will allow for powerful tests of the universality and process-independence of multi-parton correlators by comparing \pPb\ and \eAu\ data at the LHC and  EIC.

With the addition of FoCaL, ALICE will have unique capabilities at the LHC to perform photon-induced measurements as a function of rapidity gap to ensure exclusivity and to measure inclusive photo production in a wide variety of processes.
Using FoCaL and the central ALICE detectors, angular correlations can be measured to probe EPR-type signals and quantum entanglement in a kinematic domain unavailable elsewhere~\cite{Bellwied:2018gck}.
The FoCal also significantly enhances the ALICE capabilities to study the origin of long range flow-like correlations in \pp\ and \pPb\ collisions, and to quantify jet quenching effects at forward rapidity in \PbPb\ collisions.


The FoCal layout consists of a high-granularity electromagnetic calorimeter backed by a hadron calori\-meter, located outside the ALICE solenoid magnet at a distance of $7$~m from the ALICE interaction point. 
The electromagnetic part of FoCal is a compact silicon-tungsten (Si+W) sampling electromagnetic calorimeter with longitudinal segmentation. 
The sampling in the current FoCal design consists of 18 layers of tungsten and silicon pads with low granularity~($\sim $1 cm$^2$) and two~(or three) layers of tungsten and silicon pixels with high granularity~($\sim30\times30\,\mu$m$^2$).
The pad layers provide the measurement of the shower energy and profile, while the pixel layers enable two-photon separation with high spatial precision to discriminate between isolated photons and
merged showers of decay photon pairs from neutral pions. 
The hadronic part of FoCal is a conventional metal/scintillating calorimeter with high granularity of up to $2.5\times2.5$~cm$^2$, which provides good hadronic resolution and  compensation.

The proposed calorimeter will be unique in its capability to measure the inclusive direct photon distributions in \pp\ and \pPb\ collisions in the forward region for $2<\pt<20$~\GeVc.
An accuracy of 20\% is reached at $\pt \approx 4~\GeVc$ which improves to about 5\% at 10 \GeVc{} and above~(\Fig{39}), strongly constraining especially nuclear PDFs below $x\sim0.001$~\cite{AbdulKhalek:2020yuc}. 
In addition, the inclusive $\pi^0$ distribution in central \PbPb\ collisions can be measured with a systematic uncertainty below 10\% for $\pt>10$~\GeVc~(\Fig{46}), allowing for identified particle measurements at uniquely forward rapidity in \PbPb\ collisions at the LHC. 

Several prototype detectors were constructed and their performance was studied to validate the design choices for the electromagnetic part of FoCal~\cite{Muhuri:2014xva,Nooren:2017jsz,Muhuri:2019hvg,Cormier:2019rrw}. 
The results from these tests confirm the feasibility of the design concept.
For the final design, more R\&D on the integration of the system is necessary, while only modest additional R\&D is needed to finalize the pad and pixel sensor readout.
In addition, a prototype for clinical application in computer tomography based on proton tracking with a high-granularity~(pixel based) digital tracking calorimeter is being constructed by members of the FoCal collaboration~\cite{protom}.

\bibliography{focal-loi}

\providecommand{\href}[2]{#2}\begingroup\raggedright\begin{thebibliography}{10}

\bibitem{ALICECollaboration:2020rog}
{\bfseries ALICE} Collaboration, ``{Letter of Intent: A Forward Calorimeter
  (FoCal) in the ALICE experiment}'', {\em CERN-LHCC-2020-009} (6, 2020) .
  \url{https://cds.cern.ch/record/2719928}.

\bibitem{Kovarik:2015cma}
K.~Kovarik {\em et~al.}, ``{nCTEQ15 - Global analysis of nuclear parton
  distributions with uncertainties in the CTEQ framework}'',
  \href{http://dx.doi.org/10.1103/PhysRevD.93.085037}{{\em Phys. Rev.}
  {\bfseries D93} no.~8, (2016) 085037},
\href{http://arxiv.org/abs/1509.00792}{{\ttfamily arXiv:1509.00792 [hep-ph]}}.

\bibitem{Eskola:2016oht}
K.~J. Eskola, P.~Paakkinen, H.~Paukkunen, and C.~A. Salgado, ``{EPPS16: Nuclear
  parton distributions with LHC data}'',
  \href{http://dx.doi.org/10.1140/epjc/s10052-017-4725-9}{{\em Eur. Phys. J.}
  {\bfseries C77} no.~3, (2017) 163},
\href{http://arxiv.org/abs/1612.05741}{{\ttfamily arXiv:1612.05741 [hep-ph]}}.

\bibitem{AbdulKhalek:2019mzd}
{\bfseries NNPDF} Collaboration, R.~Abdul~Khalek, J.~J. Ethier, and J.~Rojo,
  ``{Nuclear parton distributions from lepton-nucleus scattering and the impact
  of an electron-ion collider}'',
  \href{http://dx.doi.org/10.1140/epjc/s10052-019-6983-1}{{\em Eur. Phys. J.}
  {\bfseries C79} no.~6, (2019) 471},
\href{http://arxiv.org/abs/1904.00018}{{\ttfamily arXiv:1904.00018 [hep-ph]}}.

\bibitem{Balitsky:1995ub}
I.~Balitsky, ``{Operator expansion for high-energy scattering}'',
  \href{http://dx.doi.org/10.1016/0550-3213(95)00638-9}{{\em Nucl. Phys. B}
  {\bfseries 463} (1996) 99--160},
  \href{http://arxiv.org/abs/hep-ph/9509348}{{\ttfamily arXiv:hep-ph/9509348}}.

\bibitem{Kovchegov:1999yj}
Y.~V. Kovchegov, ``{Small x F(2) structure function of a nucleus including
  multiple pomeron exchanges}'',
  \href{http://dx.doi.org/10.1103/PhysRevD.60.034008}{{\em Phys. Rev. D}
  {\bfseries 60} (1999) 034008},
  \href{http://arxiv.org/abs/hep-ph/9901281}{{\ttfamily arXiv:hep-ph/9901281}}.

\bibitem{Mueller:2001uk}
A.~H. Mueller, ``{A Simple derivation of the JIMWLK equation}'',
  \href{http://dx.doi.org/10.1016/S0370-2693(01)01343-0}{{\em Phys. Lett.}
  {\bfseries B523} (2001) 243--248},
\href{http://arxiv.org/abs/hep-ph/0110169}{{\ttfamily arXiv:hep-ph/0110169
  [hep-ph]}}.

\bibitem{Lappi:2016fmu}
T.~Lappi and H.~Mäntysaari, ``{Next-to-leading order Balitsky-Kovchegov
  equation with resummation}'',
  \href{http://dx.doi.org/10.1103/PhysRevD.93.094004}{{\em Phys. Rev.}
  {\bfseries D93} no.~9, (2016) 094004},
\href{http://arxiv.org/abs/1601.06598}{{\ttfamily arXiv:1601.06598 [hep-ph]}}.

\bibitem{Gribov:1984tu}
L.~Gribov, E.~Levin, and M.~Ryskin, ``{Semihard Processes in QCD}'',
  \href{http://dx.doi.org/10.1016/0370-1573(83)90022-4}{{\em Phys. Rept.}
  {\bfseries 100} (1983) 1--150}.

\bibitem{Mueller:1985wy}
A.~H. Mueller and J.-w. Qiu, ``{Gluon Recombination and Shadowing at Small
  Values of x}'', \href{http://dx.doi.org/10.1016/0550-3213(86)90164-1}{{\em
  Nucl. Phys. B} {\bfseries 268} (1986) 427--452}.

\bibitem{Ayala:1996em}
A.~Ayala, M.~Gay~Ducati, and E.~Levin, ``{QCD evolution of the gluon density in
  a nucleus}'', \href{http://dx.doi.org/10.1016/S0550-3213(97)00002-3}{{\em
  Nucl. Phys. B} {\bfseries 493} (1997) 305--353},
  \href{http://arxiv.org/abs/hep-ph/9604383}{{\ttfamily arXiv:hep-ph/9604383}}.

\bibitem{AyalaFilho:1997du}
A.~Ayala~Filho, M.~Gay~Ducati, and E.~Levin, ``{Parton densities in a
  nucleon}'', \href{http://dx.doi.org/10.1016/S0550-3213(97)00737-2}{{\em Nucl.
  Phys. B} {\bfseries 511} (1998) 355--395},
  \href{http://arxiv.org/abs/hep-ph/9706347}{{\ttfamily arXiv:hep-ph/9706347}}.

\bibitem{Kovchegov:1999ua}
Y.~V. Kovchegov, ``{Unitarization of the BFKL pomeron on a nucleus}'',
  \href{http://dx.doi.org/10.1103/PhysRevD.61.074018}{{\em Phys. Rev. D}
  {\bfseries 61} (2000) 074018},
  \href{http://arxiv.org/abs/hep-ph/9905214}{{\ttfamily arXiv:hep-ph/9905214}}.

\bibitem{Iancu:2001md}
E.~Iancu and L.~D. McLerran, ``{Saturation and universality in QCD at small
  x}'', \href{http://dx.doi.org/10.1016/S0370-2693(01)00526-3}{{\em Phys. Lett.
  B} {\bfseries 510} (2001) 145--154},
  \href{http://arxiv.org/abs/hep-ph/0103032}{{\ttfamily arXiv:hep-ph/0103032}}.

\bibitem{Citron:2018lsq}
Z.~Citron {\em et~al.}, ``{Future physics opportunities for high-density QCD at
  the LHC with heavy-ion and proton beams}'', in {\em {HL/HE-LHC Workshop:
  Workshop on the Physics of HL-LHC, and Perspectives at HE-LHC Geneva,
  Switzerland, June 18-20, 2018}}.
\newblock 2018.
\newblock
\href{http://arxiv.org/abs/1812.06772}{{\ttfamily arXiv:1812.06772 [hep-ph]}}.
\newblock

\bibitem{Benic:2016uku}
S.~Benic, K.~Fukushima, O.~Garcia-Montero, and R.~Venugopalan, ``{Probing gluon
  saturation with next-to-leading order photon production at central rapidities
  in proton-nucleus collisions}'',
  \href{http://dx.doi.org/10.1007/JHEP01(2017)115}{{\em JHEP} {\bfseries 01}
  (2017) 115}, \href{http://arxiv.org/abs/1609.09424}{{\ttfamily
  arXiv:1609.09424 [hep-ph]}}.

\bibitem{Benic:2018hvb}
S.~Beni\'c, K.~Fukushima, O.~Garcia-Montero, and R.~Venugopalan,
  ``{Constraining unintegrated gluon distributions from inclusive photon
  production in proton--proton collisions at the LHC}'',
  \href{http://dx.doi.org/10.1016/j.physletb.2019.02.007}{{\em Phys. Lett. B}
  {\bfseries 791} (2019) 11--16},
  \href{http://arxiv.org/abs/1807.03806}{{\ttfamily arXiv:1807.03806
  [hep-ph]}}.

\bibitem{Roy:2019hwr}
K.~Roy and R.~Venugopalan, ``{NLO impact factor for inclusive photon$+$dijet
  production in $e+A$ DIS at small $x$}'',
  \href{http://dx.doi.org/10.1103/PhysRevD.101.034028}{{\em Phys. Rev. D}
  {\bfseries 101} no.~3, (2020) 034028},
  \href{http://arxiv.org/abs/1911.04530}{{\ttfamily arXiv:1911.04530
  [hep-ph]}}.

\bibitem{Bellwied:2018gck}
R.~Bellwied {\em et~al.}, ``{Quantum entanglement in the initial and final
  state of relativistic heavy ion collisions}'',
  \href{http://dx.doi.org/10.1088/1742-6596/1070/1/012001}{{\em J. Phys. Conf.
  Ser.} {\bfseries 1070} no.~1, (2018) 012001},
  \href{http://arxiv.org/abs/1807.04589}{{\ttfamily arXiv:1807.04589
  [nucl-th]}}.

\bibitem{AbdulKhalek:2020yuc}
R.~Abdul~Khalek, J.~J. Ethier, J.~Rojo, and G.~van Weelden, ``{nNNPDF2.0: Quark
  Flavor Separation in Nuclei from LHC Data}'',
  \href{http://arxiv.org/abs/2006.14629}{{\ttfamily arXiv:2006.14629
  [hep-ph]}}.

\bibitem{Muhuri:2014xva}
S.~Muhuri, S.~Mukhopadhyay, V.~B. Chandratre, M.~Sukhwani, S.~Jena, S.~A. Khan,
  T.~K. Nayak, J.~Saini, and R.~N. Singaraju, ``{Test and characterization of a
  prototype silicon–tungsten electromagnetic calorimeter}'',
  \href{http://dx.doi.org/10.1016/j.nima.2014.07.019}{{\em Nucl. Instrum.
  Meth.} {\bfseries A764} (2014) 24--29},
\href{http://arxiv.org/abs/1407.5724}{{\ttfamily arXiv:1407.5724
  [physics.ins-det]}}.

\bibitem{Nooren:2017jsz}
de~Haas A.\~P.\ {\em et~al.}, ``{The FoCal prototype—an extremely
  fine-grained electromagnetic calorimeter using CMOS pixel sensors}'',
  \href{http://dx.doi.org/10.1088/1748-0221/13/01/P01014}{{\em JINST}
  {\bfseries 13} no.~01, (2018) P01014},
\href{http://arxiv.org/abs/1708.05164}{{\ttfamily arXiv:1708.05164
  [physics.ins-det]}}.

\bibitem{Muhuri:2019hvg}
S.~Muhuri {\em et~al.}, ``{Fabrication and beam test of a silicon-tungsten
  electromagnetic calorimeter}'',
  \href{http://dx.doi.org/10.1088/1748-0221/15/03/P03015}{{\em JINST}
  {\bfseries 15} no.~03, (2020) P03015},
  \href{http://arxiv.org/abs/1911.00743}{{\ttfamily arXiv:1911.00743
  [physics.ins-det]}}.

\bibitem{Cormier:2019rrw}
T.~Awes {\em et~al.}, ``{Design and Performance of a Silicon Tungsten
  Calorimeter Prototype Module and the Associated Readout}'',
\href{http://arxiv.org/abs/1912.11115}{{\ttfamily arXiv:1912.11115
  [physics.ins-det]}}.

\bibitem{protom}
H.~E.~S. Pettersen {\em et~al.}, ``{Design optimization of a pixel-based range
  telescope for proton computed tomography}'',
  \href{http://dx.doi.org/10.1016/j.ejmp.2019.05.026}{{\em Physica Medica}
  {\bfseries 63} (2019) 87--97}.

\end{thebibliography}\endgroup
\bibliographystyle{utphys}
\end{document}